\documentclass[10pt]{article} 
\usepackage{amsfonts}
\usepackage{amssymb}
\usepackage{amstext}

\usepackage{theorem}
\newtheorem{teo}{}
\newtheorem{teor}[teo]{Theorem}
\newtheorem{prop}[teo]{Proposition}
\newtheorem{coro}[teo]{Corollary}

\newtheorem{lema}[teo]{Lemma}
\newtheorem{ejem}[teo]{Example}

\newenvironment{dem}
	{\par {\it Proof:}}
 	{\hfill $\square$ \medskip}

\newcommand{\salta}{\par\vskip0.5\baselineskip}

\def\ds{\displaystyle}
\newcommand{\mc}{\mathcal}

\def\b{\begin{equation}}
\def\e{\end{equation}}
\def\be{\begin{eqnarray}}
\def\ee{\end{eqnarray}}
\def\O{\Omega}
\def\o{\omega}
\def\l{\lambda}
\def\vep{\varepsilon}
\def\ep{\epsilon}
\def\ot{\otimes}
\def\pp{\,\text{P.P.}}% Principal Part
\def\pv{{\rm P.P.\,}}
\def\<{\langle}
\def\>{\rangle}

\def\D{{\mathcal D}}
\def\H{{\mc H}}
\def\S{{\mathcal S}}
\def\N{\mathbb N}
\def\R{\mathbb R}

\date{July 2003}

\begin{document}

\title{The Role of Resonances in the Stochastic Limit\footnote{In Memory of I. Prigogine. Presented at the CFIF workshop on time asymmetric quantum theory: the theory of resonances, Lisbon, Portugal, 23--26 July, 2003.}}

\author{Fernando G\'omez$^1$}

\maketitle
\begin{center}
{\small
$^1$ Departamento de An\'alisis Matem\'atico. Universidad de Valladolid.\\
Facultad de Ciencias. Prado de la Magdalena, s.n.\\ 
47005, Valladolid, Spain.\\
e-mail: {\tt fgcubill@am.uva.es}.}
\end{center}

\begin{abstract}
In the stochastic limit the resonances play a fundamental role because they determine the generalized susceptivities which are the building blocks of all the physical information which survives in this limit.
There are two sources of possible divergences: one related to the singularities of the form factor; another to the chaoticity of the spectrum.
The situation will be illustrated starting from the example of the discrete part of the hydrogen atom in interaction with the electromagnetic field.
\end{abstract}

%\tableofcontents

%---------------------
\section{Introduction}
\label{l1}

Beyond the scheme {\it microscopic-mesoscopic-macroscopic}, there are many levels of description, probably an infinite hierarchy, in which the specific properties of a given level express some kind of {\it cumulative} or {\it collective} behaviour of properties of systems corresponding to lower levels.

These cumulative phenomena are, typically, nonlinear effects.
In absence of generally applicable methods, one introduces {\it asymptotic methods} which approximate the values of individual quantities of physical interest.
Among these asymptotic methods, the {\it scattering theory} is concerned with the {long-term behaviour} of physical system, for which $t\to\infty$, and the {\it perturbation theory} is concerned with {weak effects}, for which $\l\to 0$, where $\l$ is a parameter upon which the interaction Hamiltonian depends.

The {\it stochastic limit} \cite{ALV} puts together the scattering and perturbation theories by studying the {\it long-term cumulative effects of weak actions}, working as a {magnifying glass} of all phenomena pertaining to the scales of magnitudes we are interested in and as a {filter} of those pertaining to all the remaining scales. 

One starts from the usual quantum Hamiltonian dynamics in {\it interaction representation}, see Sect.\ref{l2}.
Depending such dynamics on a parameter $\l$, one rescales some parameters in the associated evolution (typically {\it time}) and obtains in the limit $\l\to 0$ a new dynamical system driven by a singular  Hamiltonian.
The new unitary evolution is an approximation of the original one which preserves much nontrivial information on the original complex system, see Sect.\ref{l3}.

The limit Hamiltonian is a functional of some {\it white noise}, see Sect.\ref{l5}.
The idea is that, if we look at the fast degrees of freedom of a nonlinear system with a clock, adapted to the slow ones, then the former look like an independent increment process, typically white noise.

Moreover, in the limit we shall have not a single, but an infinity of independent quantum noises, one for each Bohr frequency of the system.
This is the {\it stochastic resonance principle}, see Theorem \ref{alvt5.2.1}.

A standard scheme to describe {\it dissipation} and {\it irreversibility} passes through the so-called {\it open system} (or {\it system-reservoir}) approach, the basic physical idea of which is that dissipation an irreversible behaviour arises when two systems, traditionally denoted $S$ ({\it system}: slow degrees of freedom) and $R$ ({\it reservoir}: fast degrees), interact and one of them ($R$) exhibits some {macroscopic} or {chaotic} features, see Sect.\ref{l4}.

In the open-system approach the {\it master equation} corresponds to the adiabatic elimination of the fastly relaxing variables, a technique also called {\it coarse graining}, obtained just by taking the partial expectation of Langevin equations (stochastic limit of the Heisenberg evolution) with respect to the reference state of the master field.

The stochastic limit goes far beyond the master equation because it does not eliminate the fast degrees of freedom.
This allows to estimate the probabilities of some collective states, or more generally, the behaviour of a complex (nonlinear) system with many degrees of freedom in terms of relatively few functions of the microscopic characteristics of the quickly relaxing degrees of freedom. According to the interpretation, these functions are called {\it order parameters}, {\it kineticies}, {\it susceptibilities} or {\it susceptivities}, {\it transport coefficients}, etc, see Sect.\ref{l5}.

There is one {\it generalized susceptivity factor} for each Bohr frequency; its real part is a $\delta$-function and its imaginary part a generalized Hilbert transform, both over each {\it resonant surface}, see Eq. (\ref{alv4.21.3}).
We shall establish some connections between these real and imaginary parts in Sect.\ref{l6}.

The transport coefficients appear as {\it Ito correction terms} in the white noise Hamiltonian equations, which introduce 
quantum mechanical fluctuation-di\-ssi\-pa\-tion phenomena in the evolution.
The imaginary part of the Ito correction term corresponds to a {\it global shift} in the spectrum of the system Hamiltonian, whereas the evolution corresponding to the real part of the Ito term is a contraction, i.e. dissipative. 

Finally, in Sect.\ref{l7} we shall illustrate the situation with a concrete physical example. We will study the bound states of the hydrogen atom in interaction with the electromagnetic field.  In this case we shall obtain some sufficient conditions for the existence for the generalized susceptivity factors and calculate them explicitly.

%---------------------------------------

\section{The Interaction Representation}
\label{l2}

Let $H$ be a Hamiltonian and let $\psi(t)$ be a solution of the corresponding Schr\"odinger equation ($\hbar=1$)
$$
\partial_t \psi(t)=-iH\psi(t).
$$
Given a decomposition of $H$ into a {\it free} and an {\it interacting} parts,
$$
H=H_0+H_I,
$$
the {\it wave function $\Psi_I(t)$ in the interaction representation} is defined by
\b\label{alv1.3.2}
\psi_I(t):= e^{itH_0}\psi(t)= e^{itH_0} e^{-it(H_0+H_I)}\psi_0,
\e
and it satisfies the {\it Schr\"odinger eq. in the interaction representation}
\b\label{alv1.3.3}
\partial_t \psi_I(t)= -iH_I(t)\psi_I(t),\quad  H_I(t):=e^{itH_0}H_I e^{-itH_0}.
\e
The solution of Eq. (\ref{alv1.3.3}) with initial condition $\psi_I(t_0)$ is given by
$\psi_I(t)=U(t,t_0)\psi_I(t_0)$, where the {\it propagator} $U$ is 
\b\label{alv1.3.6}
U(t,t_0)= e^{itH_0} e^{-i(t-t_0)(H_0+H_I)} e^{-it_0H_0},
\e
which verifies
\b\label{alv1.3.7}
U(t,t_0)= U(t,s)U(s,t_0),\quad U(t,t_0)^\ast=U(t_0,t),\quad U(t,t)=I,
\e
\b\label{4a}
U(t+r,t_0+r)=e^{irH_0}U(t_1t_0)e^{-irH_0}
\e
\b\label{alv1.3.9}
\partial_t U(t,t_0)= -i H_I(t) U(t,t_0),\quad U(t_0,t_0)=I.
\e

The {\it S matrix} is formally defined as the limit
\b\label{alv1.3.10}
S=\lim_{t\to\infty} U(t,-t).
\e
 
In interaction representation the {\it Heisenberg evolution} is 
$$
j_{t,t_0}(A)=A_t:= U(t,t_0)^\ast A U(t,t_0)
$$
and satisfies the {\it flow equation}
\b\label{alv1.4.8}
j_{t,t_0}(A)=j_{s,t_0}j_{t,s}(A),
\e
whose differential form is the {\it Heisenberg equation in the interaction representation}
\b\label{alv1.4.9}
\partial_t A_t=-i[H_I(t),A_t].
\e

In what follows, by a {\it dynamical system} we mean a pair $\{\H,H\}$, where $\H$ is a Hilbert space and $H$ a Hamiltonian, or we mean a triple system 
$$
\{\H,H_0,U(t,s)\},
$$
where $\{\H,H_0\}$ is an integrable dynamical system and $U(t,s)$ is a propagator satisfying the cocycle equations (\ref{alv1.3.7}) and (\ref{4a}). 
If $U(t,s)$ is differentiable with respect to $t$, then the two definitions are equivalent.
However, the generalized definition also allows the possibility for $U(t,s)$ to satisfy a {\it stochastic differential} or {\it white noise Hamiltonian equation}.

%-------------------------------------------------
\section{A First Approach to the Stochastic Limit}
\label{l3}

The stochastic limit approximates the fundamental laws themselves in the following sense: One starts with a family of quantum dynamical systems $\{\H_\l,H_0^{(\l)},U_t^{(\l)}\}$, depending on a parameter $\l$, verifying the symbolic relation\footnote
{
In Sect.\ref{l5} we will see that the limit (\ref{alv1.6.2}) has to be understood in the sense of correlators.
}
\b\label{alv1.6.2}
\lim_{\l\to 0} \{\H_\l,H_0^{(\l)},U_{t/\l^2}^{(\l)}\}=\{\H,\tilde H_0,U_t\},
\e
where $\{\H,\tilde H_0,U_t\}$ is a new quantum dynamical system.

The equation satisfied by the dynamical evolution $U_t^{(\l)}$ is\footnote
{
Note that $U_t^{(\l)}$ is the adjoint of the {\it backward wave operator} at time $t$, i.e.
$U_t^{(\l)}=:\Omega_-^{(\l)\ast}(t) \to \Omega_-^{(\l)\ast}$, 
where $\Omega_-^{(\l)}$ is the backward wave operator.
} 
\b\label{alv1.9.1}
\partial_t U_t^{(\l)}=-i H_I^{(\l)}(t) U_t^{(\l)},\quad U_0^{(\l)}=I,
\e
interpreted as a Schr\"odinger equation in interaction picture (and not  as the usual Schr\"odinger equation with time-dependent Hamiltonian).
With the change of variables $t\mapsto \l^2t$ Eq. (\ref{alv1.9.1}) takes the form
\b\label{alv1.9.7}
\partial_t U_{t/\l^2}^{(\l)}=-i\frac{1}{\l^2} H_I^{(\l)}(t/\l^2) U_{t/\l^2}^{(\l)},\quad U_0^{(\l)}=I.
\e

The first step of the stochastic limit is to prove that
\b\label{alv1.9.8}
\lim_{\l\to 0} \frac{1}{\l^2} H_I^{(\l)}(t/\l^2)=H_t,
\e
this implies $U_{t/\l^2}^{(\l)}\to U_t$ and $U_t$ is the solution of 
$$
\partial_tU_t=-iH_tU_t.
$$

The following lemma says us that when $H_I^{(\l)}(t)$ is independent of $\l$ and an integrable function of $t$, then the limit Hamiltonian $H_t$ is simply a multiple of the $\delta$-function in $t$ characterizing the {\it white noise correlations}.

\begin{lema}\label{alvl1.9.1}
For any integrable function $F$ on $\R$ one has
\b\label{alv1.9.11}
\lim_{\l\to 0} \frac{1}{\l^2} F\left(\frac{\tau-t}{\l^2}\right)= \delta(\tau-t) \int_\R F(\sigma)\,d\sigma,
\e
in the sense that, for any bounded continuous function $\psi$ on $\R$, 
\b\label{alv1.9.12}
\lim_{\l\to 0} \int_\R \frac{1}{\l^2} F\left(\frac{\tau-t}{\l^2}\right)\,\psi(\tau)\,d\tau= 
\psi(\tau) \int_\R F(\sigma)\,d\sigma.
\e
\end{lema}

Why the rescaling $t\to t/\l^2$ as the new time scale is best understood by considering second-order perturbation theory:

\begin{lema}\label{alvl1.8.1}
Denote by $\<\cdot\>=\<\Phi,\cdot\Phi\>$ the expectation value with respect to a fixed vector $\Phi$.
Suppose that the Hamiltonian has mean zero, is time-translation invariant and such that the function $s\mapsto \<H_I(0)H_I(s)\>$ is integrable, i.e.
$$
\<H_I(t)\>=0,
$$
$$
\<H_I(t_1+s)\cdots H_I(t_n+s)\>=\<H_I(t_1)\cdots H_I(t_n)\>,
$$
$$
\int_\R |\<H_I(0)H_I(t)\>|\,dt<\infty.
$$
Then the expectation value of the second-order term of the iterated series for $U_t^{(\l)}$, i.e.
\b\label{alv1.8.5}
-\l^2\int_0^t dt_1 \int_0^{t_1} dt_2\,\<H_I(t_1)H_I(t_2)\>,
\e
has a finite nonzero limit as $\l\to 0$ and $t\to\infty$ iff
$$
\lim_{\l\to 0,\,t\to\infty} \l^2 t=\tau=c^{te}\neq 0 (<\infty).
$$
In this case the limit of (\ref{alv1.8.5}) as $\l\to 0$ is equal to 
\b\label{alv1.8.7}
-\tau \int_{-\infty}^0 ds\,\<H_I(0)H_I(s)\>.
\e
\end{lema}

The integral in Eq. (\ref{alv1.8.7}) can be interpreted as a {\it quantum transport coefficient}.
In several cases (e.g. in the weak coupling, but not in the low-density) the transport coefficient allows us to compute the lifetimes and energy shifts in agreement with the Fermi golden rule.

%---------------------
\section{Open Systems}
\label{l4}

Given two quantum dynamical systems, the {\it system} $S=\{\H_S,H_S\}$ and the {\it reservoir} $R=\{\H_R,H_R\}$, the quantum dynamical {\it composite system} will be of the form
$$
\{\H_S\ot \H_R, H_{SR}=H_S\ot 1_R+ 1_S\ot H_R+H_I\},
$$
where the {\it interaction Hamiltonian} $H_I$ contains all the new physics, with respect to the isolated systems, while $H_0=H_S+H_R$ (resp. $H_S$, $H_R$) is the {\it free Hamiltonian} (resp. of $S$, $R$).

As {\it reservoir} $R$ we will consider a quantum field $a_k,\,a_k^+$ over $\R^d$, $d\geq 3$, for which the quantities of physical interest are the {\it correlators} $\<a_{k_1}^{\ep_1}\cdots a_{k_n}^{\ep_n}\>$,
where $\<\cdot\>$ is a {\it expectation value} or {\it state} (i.e. a positive linear functional on the field algebra).
Giving an expectation value is equivalent to giving a representation of the field algebra in a Hilbert space $\H$ and a unit vector $\Phi\in\H$ such that
$$
\<a_{k_1}^{\ep_1}\cdots a_{k_n}^{\ep_n}\>=
\<\Phi,a_{k_1}^{\ep_1}\cdots a_{k_n}^{\ep_n}\Phi\>
$$

Recall that the field $a_{k}$, $a_{k'}^{+}$ together with the expectation value $\<\cdot\>$ is called a {\it mean zero Gaussian field} if 
$$
\<a_{k_1}^{\ep_1}\cdots a_{k_n}^{\ep_n}\>=0,\quad \text{if } n \text{ is odd},
$$
$$
\<a_{k_1}^{\ep_1}\cdots a_{k_{2p}}^{\ep_{2p}}\>=
\sum_{{\mc P}_0(2p)} \ep(i_1,j_1;\ldots i_p,j_p)\, \<a_{k_{i_1}}^{\ep_{i_1}} a_{k_{j_1}}^{\ep_{j_1}}\>
\cdots \<a_{k_{i_p}}^{\ep_{i_p}} a_{k_{j_p}}^{\ep_{j_p}}\>,
$$
where ${\mc P}_0(2p)$ is a subset of the ordered partitions $(i_1,j_1;\ldots i_p,j_p)$ of the set $(1,\ldots,2p)$ such that $i_1<\cdots <i_p$ and $i_\alpha<j_\alpha$, and $\ep(i_1,j_1;\ldots i_p,j_p)$ is a complex number.

For a mean zero Gaussian field the state $\<\cdot\>$ is determined by its covariance
$$
\left(
\begin{array}{cc}
\<a_{k}^{+} a_{k'}^{}\> & \<a_{k}^{} a_{k'}^{}\>
\\
\<a_{k}^{+} a_{k'}^{+}\> & \<a_{k}^{} a_{k'}^{+}\>
\end{array}
\right).
$$

The $a_{k}^{\ep}$ are operator-valued distributions.
The translation from the distribution to the operator language requires integration against suitable test functions $g$:
$$
A^+(g)=\int_M g(k)a_k^+\,dk,\quad
A(g)=\int_M \overline{g}(k)a_k\,dk.
$$
The rigorous meaning of the  multiplication of distributions $a_{k}^{\ep}$ is the multiplication of the corresponding operators $A^\ep(g)$.

In what follows we shall assume that the {\it reservoir} $R$ is a mean zero Gaussian quantum field $a_k,\,a_k^+$ over $\R^d$, $d\geq 3$, which satisfies the $q$-commutation relations 
$$
[a_k,a_{k'}^+]_q=a_{k}a_{k'}^{+}- qa_{k'}^{+}a_{k}=\delta(k-k')
$$
for some complex number $q$.
In this case $a_{k}$ are called {\it annihilators} and $a_{k'}^{+}$ {\it creators}.
We speak of a {\it boson field} if $q=1$, of a {\it Fermi field} if $q=-1$, of a {\it Boltzmann field} if $q=0$.

If $a_k, a_k^+$ are boson creation and annihilation operators, there exists a real-valued function $\o(k)$ such that the associated {\it free Hamiltonian} has the form
\b\label{alv2.13.2}
H_0 = \int \o(k)a^+(k)a(k)\,dk,
\e
in the sense that the commutator with $H_0$ of any polynomial in the field operators coincides with the commutator of the same polynomial with the right hand side of Eq. (\ref{alv2.13.2}).
In particular, for all $k$ we have
$$
\partial_t a_k(t)=-i[H_0,a_k(t)]=-i\o_k a_k(t)
\quad \Rightarrow \quad
a_k(t)=e^{-it\o_k} a_k(0).
$$
The function $\o(k)$ is called the {\it free 1-particle Hamiltonian}.
Usually one wants $\o(k)$ to be positive and vanishes nowhere.
For example, $k^2/(2m)$, $\sqrt{k^2+m^2}$, $|k|$.
\salta

Consider a system-reservoir Hamiltonian 
$$
H_{SR}=H_S\ot 1_R+ 1_S\ot H_R+H_I,
$$
for which the interaction Hamiltonian $H_I$ 
is of {\it dipole-type}, i.e. it has the form
\b\label{alv4.8.1}
H_I=\int dk\, \big\{ D(k)\ot a^+(k)+D^+(k)\ot a(k)\big\},
\e
where $\{D(k):k\in\R^d\}$ is a family of system operators (acting on $\H_S$) called the {\it response terms} or {\it currents} and containing local information about the interaction.

If the free system Hamiltonian $H_S$ has a discrete spectrum, i.e.
$$
H_S=\sum_n \vep_n P_n=\sum_n \vep_n |\vep_n\>\<\vep_n|,
$$
the time evolved interaction Hamiltonian becomes
\b\label{alv4.8.3}
H_I(t)=\int dk\,\sum_{m,n} P_m D(k) P_n\ot e^{it(\o_k + \vep_m-\vep_n)} a_k^+ +h.c.
\e
Introducing the operators 
\b\label{opD}
D_\o(k):=\sum_{\vep_n-\vep_m=\o} P_m D(k) P_n= \sum_{\vep_n-\vep_m=\o} \<\vep_n|D(k)|\vep_m\>, |\vep_m\>\<\vep_n|,
\e
for which we have $e^{itH_S} D_\o e^{-itH_S}= e^{-it\o} D_\o(k)$, Eq. (\ref{alv4.8.3}) becomes
\b\label{alv4.8.7}
H_I(t)=\int dk\,\sum_{\o} D_\o(k)\ot e^{it(\o_k -\o)} a_k^+ +h.c.
\e

Moreover, assume that the system operators $D_\o(k)$ verify the {\it generalized dipole approximation}, i.e. 
$$
D_\o(k)=D_\o^{dipole}(k)=g(k)D_\o,
$$
 where $g$ is a test function, the {\it cutoff} or {\it form factor} describing the strength of the interaction of the system with the environment, and $D_\o$ is a fixed system operator. Then, if there is only 
{\it one Bohr frequency} $\o$, the interaction Hamiltonian (\ref{alv4.8.1}) becomes
$$
H_I=D^+\ot A(g)+D\ot A^+(g)=\int_{\R^d} dk\, \{D^+\ot \overline{g}(k)a_k+D\ot g(k)a^+_k\}.
$$
and its free evolution is
\b\label{alv4.11.1}
\begin{array}{rl}
H_I(t) &
\ds =\int dk\, \left\{D^+\ot\overline{g}_k e^{-it(\o_k-\o)}a_k + D\ot g_k e^{it(\o_k-\o)}a_k^+\right\}
\\[2ex]
& \ds =: D^+a_t+Da_t^+,
\end{array}
\e
where we omit the symbol $\ot$ and 
$$
a_t:=\int dk\,\overline{g}_k e^{-it(\o_k-\o)}a_k.
$$

Let us show how quantum white noises arise as stochastic limits of free fields.

%-------------------------------------------
\section{The Stochastic Resonance Principle}
\label{l5}

A quantum field $b^\pm(t,k)$ ($t\in\R,\,k\in\R^d$) with expectation value $\<\cdot\>$ is called a {\it white noise} if it is a mean zero Gaussian field with covariance
$$
\< b^\ep(t,k) b^{\ep'}(t',k')\>=\delta(t-t')G_{\ep,\ep'}(k,k'),
$$
where $G_{\ep,\ep'}(k,k')$ is a positive definite distribution (as a function of the variables $\ep, k,\ep',k'$).

The convergence will be taken in the following sense.
Let $a^\pm_\l(k)$ be a family of fields with cyclic vector $\Phi_\l$, parametrized by a real number $\l$, and let $a^\pm_k$ be another field with cyclic vector $\Phi$.
$a^\pm_\l(k)$ is said to {\it converge} to $a^\pm_k$ {\it in the sense of correlators}\footnote
{
The correlation functions are also called {\it Wightman functions} in quantum field theory or {\it mixed moments} in probability theory.
}
as $\l\to\l_0$ if, for any natural $n$, for any $k_1\ldots k_n\in\R^d$, and any choice of $\ep_1\ldots \ep_n\in \{0,1\}$, one has, in the sense of distributions,
$$
\lim_{\l\to\l_0} \<\Phi_\l, a^{\ep_1}_\l(k_1)\cdots a^{\ep_n}_\l(k_n)\Phi_\l\>=
\<\Phi, a^{\ep_1}_{k_1}\cdots a^{\ep_n}_{k_n}\Phi\>.
$$
It is well known that if, as $\l\to 0$, the correlators $\<\Phi_\l, a^{\ep_1}_\l(k_1)\cdots a^{\ep_n}_\l(k_n)\Phi_\l\>$ converge to a distribution $F(k_1\ldots k_n)$, then there exists a field $\{a^\pm_k,\H,\Phi\}$ such that $F(k_1\ldots k_n)=\<\Phi, a^{\ep_1}_{k_1}\cdots a^{\ep_n}_{k_n}\Phi\>$.
\salta

Coming back to the time evolved interaction Hamiltonian given in Eq. (\ref{alv4.11.1}).
The Schr\"odinger equation in interaction representation takes the form, see Eq. (\ref{alv1.9.1}):
\b\label{alv4.11.5}
\partial_t U_t^{(\l)}=-i H_I^{(\l)}(t) U_t^{(\l)}=-i\l\big(D^+a_t+Da_t^+\big) U_t^{(\l)}.
\e
The rescaled evolution operator satisfies the {\it rescaled Schr\"odinger equation}, see Eq. (\ref{alv1.9.7}):
\b\label{alv4.11.6}
\begin{array}{rl}
\ds \partial_t U_{t/\l^2}^{(\l)} &
\ds =-i\frac{1}{\l^2} H_I^{(\l)}(t/\l^2) U_{t/\l^2}^{(\l)}
\\[2ex]
& \ds =-i\frac{1}{\l} \big(D^+a_{t/\l^2}+Da_{t/\l^2}^+\big) U_{t/\l^2}^{(\l)}
\end{array}
\e

The next result shows that the rescaled fields $\frac{1}{\l}a_{t/\l^2}$ converge in the sense of correlators to a quantum white noise $b_t$. 
The new fields operators $b$ act on some new Hilbert space and are called the {\it master fields} or {\it quantum noises}. 

\begin{teor}\label{alvt3.2.2}
If the field $a^\pm(k)$ is mean zero Gaussian and satisfies the $q$-deformed relations
$$
a(k)a^+(k')-qa^+(k')a(k)=\delta(k-k'),
$$
then the rescaled field
$$
\frac{1}{\l}\,a^\pm(t/\l^2,k):=\frac{1}{\l}\,e^{\pm i\o(k)t/\l^2} a^\pm(k)
$$
converges in the sense of distribution correlators to a $q$-deformed white noise $b^\pm_t(k)$, i.e., satisfying
$$
b_t(k)b^+_{t'}(k')-qb^+_{t'}(k')b_t(k)=2\pi\,\delta(t-t')\,\delta(\o(k))\,\delta(k-k').
$$
\end{teor}

Thus, from Eq. (\ref{alv4.11.6}),
\b\label{alv4.11.10}
\lim_{\l\to 0} \frac{1}{\l^2} H_I^{(\l)}(t/\l^2)=H_t=D^+ b_t+Db_t^+
\e
and the {\it white noise Hamiltonian equation} is 
\b\label{alv4.11.11}
\partial_t U_t=-iH_tU_t=-i(D^+ b_t+Db_t^+)U_t.
\e

One has to bring equation (\ref{alv4.11.11}) to its {\it normally ordered form}, i.e. instead of the term $b_tU_t$ we would like to have the term $U_tb_t$. 
If the reference state of the $a$ field is the Fock vacuum, then the normally ordered form of the white noise Hamiltonian (\ref{alv4.11.11}) is the following equation
\b\label{alv4.12.6}
\partial_t U_t=-i\big(D^+U_tb_t+ Db_t^+U_t\big)-YU_t,
\e
where 
\b\label{alv4.12.7}
Y:=\gamma_-D^+D,
\e
\b\label{alv4.12.5}
\gamma_-:=\int_{-\infty}^0 dt \int_{\R^d} dk\, e^{-it(\o_k-\o)} |g(k)|^2.
\e

The operator $Y$ defined by Eq. (\ref{alv4.12.7}), the coefficient of $U_t$ in the term not including noise operators in  Eq. (\ref{alv4.12.6}), is called the {\it operator transport coefficient}, {\it drift coefficient} or {\it Ito correction term}.
This term corresponds to a non-selfadjoint correction to the system Hamiltonian that can be considered as the prototype of the {\it quantum mechanical fluctuation-dissipation relation}:
$$
iH_S\to iH_S-\gamma_-D^+D=i(H_S+Im(\gamma_-)D^+D) - Re(\gamma_-) D^+D.
$$

Since $H_S$ commutes separately with the real and imaginary part, the imaginary part of the Ito correction term is a {\it global shift} in the spectrum of the system Hamiltonian.
On the other hand, the evolution generated by $- Re(\gamma_-) D^+D$ is a contraction, in general nonunitary, i.e. {\it dissipative}. In fact, $2Re(\gamma_-)=\<b_tb_t^+\>$ is a measure of the strength of the fluctuations.

For the {\it generalized susceptivity factor} $\gamma_-$ defined by Eq. (\ref{alv4.12.5}),
using the identity 
$$
\int_{-\infty}^0 e^{-it\o}\,dt=\frac{-i}{\o-i0}=\pi\delta(\o)-i \pp \frac{1}{\o},
$$
we can write
\b\label{alv4.21.3}
\begin{array}{rl}
\gamma_- 
& \ds =\pi\< g,\delta(\o_k-\o)g\>-i\<g, \pp \frac{1}{\o_k-\o}g\>
\\[2ex]
& \ds =\pi\int_{\R^d} dk\,|g(k)|^2\delta(\o_k-\o) - i\pp \int_{\R^d} dk\,\frac{|g(k)|^2}{\o_k-\o}.
\end{array}
\e

Assuming that $\o(k)>0$ almost everywhere, the argument of the $\delta$-function in $Re(\gamma_-)$ can be zero only if $\o>0$. Since $2Re(\gamma_-)=\<b_tb_t^+\>$, the fact that the contribution relative to a given frequency vanishes means there is no master field with that characteristic frequency. So, in the stochastic limit only the master fields corresponding to positive Bohr frequencies survive.
But Eq. (\ref{alv4.21.3}) also shows that this is not the case for the imaginary part. In the stochastic limit the negative Bohr frequencies contribute with an overall red shift to the energy.
\salta

For more than one Bohr frequency the time evolved interaction Hamiltonian is of the form, see Eq. (\ref{alv4.8.7}):
\b\label{alv4.8.7r}
H_I(t)= \int dk\, \sum_{\o} D_\o g(k) e^{it(\o_k -\o)} a_k^+ +h.c.
\e
This decomposition of the interaction Hamiltonian suggests that, before the limit, the original field splits into a family of effective fields, each of which is interacting with its own Bohr frequency $\o$ and with the other effective fields.

The following theorem says that, if it is possible to interchange the sum, the integral and the limit, the mutual interaction of these effective fields becomes negligible in the stochastic limit, and after the limit this mutual independence becomes exact.

\begin{teor}\label{alvt5.2.1}
Let $a(t,k),\, a^+(t,k)$ be a mean zero Gaussian quantum field with respect to a given state $\<\cdot\>$. suppose that the process $a^\pm (t,k)$ is stationary and its covariance matrix is time-integrable in the sense of distributions, i.e.
$$
\int_\R dt\,\left| \int dkdk'\,\overline{f}(k) g(k')\<a^\ep(0,k)a^{\ep'}(t,k')\>\right|<\infty,
\quad f,g\in\S(\R^d).
$$ 
For each real number $\o$ define the new process
$$
a_\o(t,k):=e^{-it\o} a(t,k).
$$
Then the limit, in the sense of distribution correlators,
$$
\lim_{\l\to 0} \frac{1}{\l}\, a_\o(t/\l^2,k)= b_\o(t,k),
$$
exists and is the white noise $b_\o(t,k)$ of the same Gaussian type as $a(t,k)$ and with covariance
$$
\<b_\o^\ep(t,k)b_\o^{\ep'}(t',k')\>=\delta(t-t') \int_\R ds\, e^{-is\o}\<a^\ep(0,k)a^{\ep'}(s,k')\>.
$$
Moreover, the quantum white noises $\{b_\o(t,k):\o\in\R\}$ are mutually independent.
\end{teor}

This illustrates the {\it stochastic resonance principle}:  In the limit we shall have not a single, but an infinity of independent quantum noises, one for each Bohr frequency of the system.

Thus, for each Bohr frequency we will have a generalized susceptivity factor given by  Eq. (\ref{alv4.21.3}), whose real part is a $\delta$-function and its imaginary part is a generalized Hilbert transform, both over each resonant surface. 
Let us pass to study these real and imaginary parts.

%-------------------------------------------
\section{The Distribution $(\o(k)-\o)^{-1}$}
\label{l6}

Let
$$
k\in\R^d \to \o(k)-\o \in \R
$$
be a $C^\infty$-function except perhaps for some closed set of singular points with $d$-dimensional Lebesgue measure zero. In this section we study, for each test function $\phi\in\D(\R^d)$, the integral 
$$
\<\frac{1}{\o(k)-\o},\phi(k)\>=\int_{\R^d} \frac{\phi(k)}{\o(k)-\o}\, dk
$$
and its Cauchy principal value or principal part
$$
\<\pv\frac{1}{\o(k)-\o},\phi(k)\>=
\lim_{\epsilon\to 0} \int_{|\o(k)-\o|>\epsilon} \frac{\phi(k)}{\o(k)-\o}\, dk\,,
$$
just the imaginary part of the generalized susceptivity factor $\gamma_-$, see Eq.(\ref{alv4.21.3}).

To this end, assume that $\nabla \o(k) \neq 0$ for each regular point $k\in \R^d$.
Then the equation $\o(k)-\o=c^{te}$ defines a regular surface $S_{c}$ of dimension $d-1$ in a neighbourhood sufficiently small of each regular point verifying the equation and, in such neighbourhood, we can consider a local change of variables $\psi:(u_1,\ldots,u_d)\to (k_1,\ldots,k_d)$ such that the surface $S_{c}$ is given by fixing one of the $u$'s to the value of the constant $c$.  
 So, for example, we can choose $\o(k)-\o=u_1$ and arbitrarily the $u_2,\ldots,u_d$, but with the condition that the Jacobian $J\psi(u)$ is different from zero.
Then, by the change of variables theorem, we will have\footnote{
We can assume that the support of $\phi$ is contained in the neighbourhood where the local change of variables is given, else we consider a suitable locally finite partition of the unity.}
$$
\int_{\R^d} \frac{\phi(k)}{\o(k)-\o}\, dk =
\int_{\psi^{-1}(\R^d)} \frac{\phi(\psi(u))}{u_1}\,J\psi(u)\, du \,,
$$
being the last integral, by Fubini's theorem, equal to
\b\label{0}
\int_a^b \frac{1}{u_1}\left[ \int_{\psi^{-1}_{u_1}} \phi(\psi(u))\,J\psi(u)\, du_2\ldots du_d\right] \,du_1 \,,
\e
where $\psi^{-1}(\R^d)=\{u_1\times\psi^{-1}_{u_1}\,|\,u_1\in [a,b]\}$.

We can write the integral (\ref{0}) in terms of differential forms.
For it, consider the form $\O_{c}$ of order $d-1$ associated to the function $W(k)=\o(k)-\o$ on the surface $S_{c}$ by the equation
\b\label{1}
dW\wedge \O_{c}=dv\,,
\e
where $dv=dk_1\wedge\cdots\wedge dk_d$ is the volume element in $\R^d$.
Such form $\O_{c}$ exists in a certain $d$-dimensional domain containing the surface $S_{c}$ because in a neighbourhood of any point of that surface one can introduce a local system of coordinates $u_1,\ldots,u_d$ such that one of these coordinates, for example $u_j$, is the quantity $W(k)$; then, if $\psi:(u_1,\ldots,u_d)\to (k_1,\ldots,k_d)$ is the corresponding change of coordinates, we will have
$$
dv= J\psi(u)\,du_1\wedge\cdots \wedge du_{j-1} \wedge dW \wedge du_{j+1}\wedge \cdots \wedge du_d
$$
and then
\b\label{2}
\O_{c} = (-1)^{j-1} J\psi\big|_{u_j=c} \, du_1\wedge\cdots \wedge du_{j-1} \wedge du_{j+1}\wedge \cdots \wedge du_d.
\e
In particular, if in a neighbourhood of a given point we have $\partial W(k)/\partial k_j\neq 0$, we can take as coordinates $u$
$$
u_1=k_1, \ldots, u_j=W,\ldots , u_d=k_d \,,
$$
we have then
$$
J\psi(u)=\frac{1}{J\psi^{-1}(k)}= 
\frac{1}{\partial W(k)/\partial k_j}
$$
and the form $\O_{c}$ defined in Eq. (\ref{2}) becomes
$$
\O_{c}= (-1)^{j-1} \frac{ dk_1\wedge\cdots \wedge dk_{j-1} \wedge dk_{j+1}\wedge \cdots 
\wedge dk_d}{\partial W/\partial k_j} \,.
$$

The form $\O_{c}$ verifying Eq. (\ref{1}) is not unique since we can add to it any form $\Lambda$ orthogonal to $dW$, that is, such that $dW\wedge \Lambda=0$.
Such forms $\Lambda$ can be written as $\Lambda=\gamma\wedge dW$, where $\gamma$ is a certain form of order $d-2$.

We note also that the form $\O_{c}$ does not depend on the choice of coordinates 
$u_1,\ldots du_{j-1}, u_{j+1},\ldots,u_d$, but it does on the function $W$ defining the surface $S_{c}$. For example, if instead of $W(k)=c$ one considers the equation $\alpha(k)\big[W(k)-c\big]=0$, where $\alpha$ is a nowhere zero function, then $d(\alpha( W-c))=\alpha\,dW+(W-c)\,d\alpha$ and along the surface $S_{c}$ we will have
$$
\O_c^1=\frac{dv}{d(\alpha(W-c))}=\frac{1}{\alpha}\O_c \,.
$$

Coming back to the situation of formula (\ref{0}), for coordinates $u_1=W, u_2,\ldots, u_d$, by formula (\ref{2}) we have
$$
\O_c=\O_{u_1}= J\psi(u)\big|_{u_1=c}\, du_2\wedge\cdots \wedge du_d
$$
and from Eq. (\ref{0}) we obtain 
\b\label{4}
\int_{\R^d} \frac{\phi(k)}{\o(k)-\o}\, dk =
\int_a^b \frac{1}{u_1} \left[ \int_{S_{u_1}} \phi(\psi(u))\,\O_{u_1}\right]\, du_1 \,.
\e

In terms of the distribution $\delta(W-u_1)=\delta(\o(k)-\o-u_1)$  formula (\ref{4}) can also be written as\footnote{
For a definition of the distribution $\delta(W-u_1)$ see, for example, the chapter 3 of \cite{GS}.} 
$$
\int_a^b \frac{1}{u_1} \<\delta(W-u_1),\phi\>\, du_1 \,.
$$

We note that the function
\b\label{7}
\Phi(u_1)=\int_{S_{u_1}} \phi(\psi(u))\,\O_{u_1}\,,
\e
that appears in formula (\ref{4}) is a test function belonging to $\D(\R)$. Indeed, the change of variables $\psi$, being bicontinuous, transforms compact sets into compact sets, and vice versa. Moreover, since $\psi$ is a $C^\infty$-function, $\phi\circ\psi\in \D(\R^d)$ and then its product with the $C^\infty$-function $J\psi$ is also in  $\D(\R^d)$. Thus, the integral with respect to $\O_{u_1}$ of that product extends  into a set of finite measure and then is a bounded $C^\infty$-function of $u_1$ with compact support. Finally, similar arguments can be applied to its derivatives.

The same reasoning can be applied when, instead of $\D(\R^d)$, we consider the Schwartz space $\S(\R^d)$, whenever the Jacobian $J\psi$ and its derivatives are of polynomial growth. 

Since $\Phi(u_1)\in\D(\R)$, the integral 
$$
\int_{\R^d} \frac{\phi(k)}{\o(k)-\o}\, dk =
\int_a^b \frac{1}{u_1}\, \Phi(u_1) \, du_1 \,.
$$
is improper only at $u_1=0$ when $a\leq 0\leq b$.
Then, for the study of the convergence of this integral it is convenient to obtain an asymptotic development of $\Phi(u_1)$ for small values of $u_1$.
To this end, let us consider the functional depending on the complex parameter $\l$ 
\b\label{6}
\<W_+^\l,\phi\>= \int_{W>0} W^\l(k)\,\phi(k)\,dk\,.
\e
If the $C^\infty$-function $W$ is such that the equation $W(k)=0$ defines locally a $d-1$ dimensional surface of regular points, that is, for each point $k_0$ in the surface there exist a neighbourhood $V$ of $k_0$ in $\R^d$ and a local system of coordinates $u_1,\ldots,u_d$ such that $W(k)=u_1$ for every $k\in V$ (for example, when $\nabla W(k_0)\neq 0$), and the same is valid for the equation $W(k)=c$ with $c>0$, then the distribution $W_+^\l$ is meromorphic with singularities the sequence of simple poles \cite[sect.3.4.2]{GS}
$$
\l= -1,-2,\ldots,-n,\ldots
$$
The residue  of the function (\ref{6}) at each of these poles can be expressed by means of the test function $\Phi(u_1)$ defined in formula (\ref{7}),
being the residue at $\l=-n$ equal to
$$
{\rm Res}(\<W_+^\l,\phi\>,\l=-n)=\frac{\Phi^{(n-1)}(0)}{(n-1)!}\,.
$$

In terms of the distributions $\delta^{(k)}(W)$, we have then
$$
{\rm Res}(\<W_+^\l,\phi\>,\l=-n)=\frac{(-1)^{n-1}}{(n-1)!}\,\<\delta^{(n-1)}(W),\phi\>
$$
and we can say that the residue of $W^\l$ at the simple pole $\l=-n$ is 
$$
{\rm Res}(W_+^\l,\l=-n)=\frac{(-1)^{n-1}}{(n-1)!}\,\delta^{(n-1)}(W)\,,
$$
in complete analogy with the unidimensional case for the distribution $x_+^\l$.

Since the behaviour of $\Phi(u_1)$ for $u_1>\epsilon>0$ has not influence over the singularities of the integral 
$$
\int_0^\infty u_1^\l\,\Phi(u_1)\,du_1\,,
$$
the knowledge of these singularities permits us write an asymptotic development of $\Phi(u_1)$ for small values of $u_1$.
Indeed, in our case \cite[sect.3.4.5]{GS}
\b\label{8}
\Phi(u_1)\simeq \sum_{n=0}^{\infty} \frac{(-1)^{n}}{n!}\,\<\delta^{(n)}(W),\phi\>\, u_1^{n},
\quad {\rm for}\,\,u_1\,\,{\rm small}\,.
\e

From these results it is easy to derive the following

\begin{prop}\label{prop1}
Let $W(k)=\o(k)-\o$ be a $C^\infty$-function, except perhaps for some closed set $E$ of singular points with $d$-dimensional Lebesgue measure zero, such that the equation $W(k)=0$ defines a $d-1$ dimensional surface $S_0$ of regular points, that is, for each point $k_0$ in the surface there exist a neighbourhood $V$ of $k_0$ in $\R^d$ and a local system of coordinates $u_1,\ldots,u_d$ such that $W(k)=u_1$ for every $k\in V$ (for example, when $\nabla W(k_0)\neq 0$), and the same is valid for the equation $W(k)=c$ with $c\in (a,b)$, where $\R^d\backslash E=\cup_{c\in(a,b)} S_c$.
Then, given a test function $\phi\in\D(\R^d)$, we have:
\begin{itemize}

\item[(i)]
When $a\leq 0\leq b$, the integral 
$$
\<\frac{1}{\o(k)-\o},\phi(k)\>=\int_{\R^d} \frac{\phi(k)}{\o(k)-\o}\, dk =
\int_a^b \frac{1}{u_1} \Phi(u_1)\, du_1 
$$
is finite if and only if $\<\delta(\o(k)-\o),\phi(k)\>=0$.

\item[(ii)]
When $a\leq 0\leq b$, the Cauchy principal value 
$$
\begin{array}{rl}
\ds \<\pv\frac{1}{\o(k)-\o},\phi(k)\> &
\ds = \lim_{\epsilon\to 0} \int_{|\o(k)-\o|>\epsilon} \frac{\phi(k)}{\o(k)-\o}\, dk 
\\[2ex]
& \ds = \lim_{\epsilon\to 0} \left[ \int_a^{-\epsilon} \frac{1}{u_1} \Phi(u_1)\, du_1
+ \int_\epsilon^b \frac{1}{u_1} \Phi(u_1)\, du_1 \right]
\end{array}
$$
is always finite.

\end{itemize} 
(Recall that, when $0\not\in [a,b]$, the integrals in (i) and (ii) are always finite.)
\end{prop}
 
\begin{dem}
Since $\Phi(u_1)\in\D(\R)$, the asymptotic development (\ref{8}) of $\Phi(u_1)$ is valid in a neighbourhood of $u_1=0$ and also we can assume that $a$ and $b$ are finite.

(i). The unidimensional integral $\int_a^b \Phi(u_1)/u_1\, du_1$ converges if and only if  
for any $\alpha>0$ we have $\Phi(u_1,\o)\simeq u_1^\alpha$ as $u_1\to 0$. By formula (\ref{8}), this condition is satisfied if and only if $\<\delta(\o(k)-\o),\phi(k)\>=0$.

(ii). It is well known \cite[theor.1.35]{SCH} that for a function of the form $\Phi(u_1)/u_1$, $\Phi$ being continuous in a neighbourhood of $u_1=0$, the integral in the $\pv$ sense exists.
\end{dem}

\begin{coro}\label{coro2}
Proposition \ref{prop1} is satisfied also for every $\phi\in\S(\R^d)$ if, in addition, the Jacobian $J\psi$ and its derivatives are of polynomial growth. 
\end{coro}

\begin{dem}
Recall that, if $\phi\in\S(\R^d)$, then also $\Phi\S(\R^d)$ when the Jacobian $J\psi$ and its derivatives are of polynomial growth. 
\end{dem}

\begin{ejem}\label{ex1}
\rm
For the radiative dispersion $\o(k)=|k|=\left(\sum_{j=1}^d k_j^2\right)^{1/2}$, if we take
\b\label{chv}
u_1=\o(k)-\o,\, u_2=\theta_1,\,\ldots,\, u_d=\theta_{d-1},
\e
where $\theta_1,\ldots,\theta_{d-1}$ are the usual angles in spherical coordinates, we obtain
$$
\O_{u_1}=d\sigma_{S_{u_1+\o}}\,,
$$
being $d\sigma_{S_{u_1+\o}}$ the Euclidean element of surface for the sphere $S_{u_1+\o}$ with centre the origin and radius $u_1+\o$. In this case we have
\b\label{iex1}
\int_{\R^d} \frac{\phi(k)}{\o(k)-\o}\, dk =
\int_{-\o}^\infty \frac{1}{u_1} \left[ \int_{S_{u_1+\o}} \phi(\psi(u))\,d\sigma_{S_{u_1+\o}}\right]\, du_1 \,.
\e

Let us put $\Phi(u_1,\o)=\int_{S_{u_1+\o}} \phi(\psi(u))\,d\sigma_{S_{u_1+\o}}$.

For $\o<0$ the integrals of Eq. (\ref{iex1}) converge because $\Phi(u_1,\o)$ is in $\D(\R)$ as function of $u_1$.

If $\o> 0$, we are under the hypothesis of proposition \ref{prop1} and then the integrals in Eq. (\ref{iex1}) converge if and only if $\<\delta(\o(k)-\o),\phi(k)\>=0$, and they converge in the $\pv$ sense for all $\phi\in\D(\R^d)$.

If $\o=0$, we cannot apply proposition \ref{prop1} since the equation $W(k)=\o(k)=0$ does not define a regular surface else a singular point $k=0$, but in this case the integrals of Eq. (\ref{iex1}) converge because $\Phi(u_1,0)\to 0$ as $u_1\to 0$.

By corollary \ref{coro2}, the same results are satisfied for $\phi\in\S(\R^d)$.
\end{ejem}

\begin{ejem}\label{ex2}
\rm
For $\o(k)=k^2=\sum_{j=1}^d k_j^2$, applying again the change of variables (\ref{chv}), now we have 
$$
\O_{u_1}=\frac{1}{2\left(\sum_{j=1}^d k_j^2\right)^{1/2}}d\sigma_{S_{\sqrt{u_1+\o}}}
=\frac{1}{2\sqrt{u_1+\o}}d\sigma_{S_{\sqrt{u_1+\o}}}\,.
$$
Then, in this case we obtain
\b\label{iex2}
\int_{\R^d} \frac{\phi(k)}{\o(k)-\o}\, dk =
\int_{-\o}^\infty \frac{1}{u_1} \left[ \int_{S_{\sqrt{u_1+\o}}} \frac{\phi(\psi(u))}{2\sqrt{u_1+\o}}\,d\sigma_{S_{\sqrt{u_1+\o}}}\right]\, du_1 \,.
\e

Here, $\Phi(u_1,\o)=\int_{S_{\sqrt{u_1+\o}}} \frac{\phi(\psi(u))}{2\sqrt{u_1+\o}}\,d\sigma_{S_{\sqrt{u_1+\o}}}$
and we can apply the same arguments than in example \ref{ex1} to determine the convergence of the integrals in (\ref{iex2}).
Thus, for $\o\leq 0$ the integrals of (\ref{iex2}) converge for every $\phi\in\D(\R^d)$ and, on the other hand, when $\o> 0$ these integrals converge if and only if $\<\delta(\o(k)-\o),\phi(k)\>=0$ and they converge in the $\pv$ sense for all $\phi\in\D(\R^d)$.

By corollary \ref{coro2}, the same results are satisfied for $\phi\in\S(\R^d)$.

\end{ejem}

These results are of applicability in the following physical example. 

%------------------------------------------
\section{The Hydrogen Atom in the EM Field}
\label{l7}

It is well known \cite{PW} that in a central potential, caused here by the hydrogen nucleus, the bound states of a spinless electron  are determined by three quantum numbers $n$, $l$ and $m$.

The {\it total} or {\it energy} quantum number $n$, whose range of values is $n=1,2,3,\ldots,+\infty$, determines the energy $E_n$ of the electron
\b\label{21}
E_n= -\frac{1}{2} \frac{Zme^4}{n^2 \hbar^2} = -\frac{Ze^2}{2a_0 n^2}\,,
\e
where $m$ and $e$ are the mass and charge of the electron, $a_0$ is the {\it Bohr radius}
\b\label{22}
a_0=\frac{\hbar^2}{me^2}=0.53\times 10^{-8}\,{\rm cm}
\e
and, for the hydrogen atom, $Z=1$.  

The {\it orbital} and {\it magnetic} quantum numbers, $l$ and $m$, determine the angular momentum and the angular momentum along the axis of quantization, respectively, and their ranges of values are $l=0,1,2,\ldots,n-1$ and $m=-l,-l+1,\ldots,+l$.

In the following $q$ shall denote the position of the electron in the 3-dimensional space and we shall assume that the nucleus of the hydrogen atom is fixed at the origin.

In spherical coordinates, the associated total eigenfunctions $\psi_{nlm}$  are given by
\b\label{23}
\psi_{nlm} (|q|,\theta,\phi)= R_{nl}(|q|)\, Y_{lm}(\theta,\phi)\,,
\e
being the radial eigenfunction  corresponding to the quantum numbers $n$ and $l$, $R_{nl}$, equal to
\b\label{24}
R_{nl}(|q|)= -\left[ \left( \frac{2}{na_0}\right)^3 
\frac{(n-l-1)!}{2n[(n+l)!]^3}\right]^{1/2}\,
e^{-|q|/2}\, |q|^l\, L_{n+l}^{2l+1}(|q|)\,,
\e
where $L_{n+l}^{2l+1}$ is the associated Laguerre polynomial
\b\label{25}
L_{n+l}^{2l+1}(|q|)= \sum_{s=0}^{n-l-1} 
\frac{ (-1)^{s+2l+1} \, [(n+l)!]^2\,|q|^s}{(n-l-1-s)!\,(2l+1+s)!\,s!}\,
\e  
and the spherical harmonics  of order $l$, $Y_{lm}$, are given by
\b\label{26}
\begin{array}{rl}
\ds Y_{lm}(\theta,\phi)= 
& \ds (-1)^m\, \left[ \frac{2l+1}{4\pi} 
\frac{(l-m)!}{(l+m)!} \right]^{1/2}\, P_l^m\big(\cos(\theta)\big)\, e^{im\phi}
\\[2ex]
= &
\ds (-1)^{l+m}\, \frac{1}{2^l\,l!}\, \left[ \frac{(2l+1)!}{4\pi} \frac{(l-m)!}{(l+m)!} \right]^{1/2}\,
\sin^{|m|}(\theta)
\\[2ex]
& \ds \times\, \left( \frac{d}{d(\cos(\theta))}\right)^{l+m} \sin^{2l}(\theta)\, e^{im\phi}\,,
\end{array}
\e
where $P_l^m$ denotes the associated Legendre function of the first kind
\b\label{27}
P_l^m(\xi)= (1-\xi^2)^{m/2}\,\frac{1}{2^l\,l!}\, \frac{d^{l+m}}{d\xi^{l+m}}(\xi^2-1)^l\,.
\e

In the following we shall consider only states of the electron with orbital number $l=0$.
Recall that $Y_{00}(\theta,\phi)=(4\pi)^{-1/2}$.
\salta

For Quantum Electrodynamics, the expansion of the EM vector potential in Fourier integral is
$$
A(q)=\int dk\, g(k)e^{ik\cdot q} a_k,
$$
where, as until now, $q$ denotes the position of the electron in the 3-dimensional space, and $k\in\R^3$ corresponds to momentum coordinates.
From this expansion we obtain the response terms
\b\label{rtqed}
D(k)=\frac{e^{ik\cdot q}}{|k|^{1/2}},
\e
being the cutoff or form factor $g(k)=|k|^{-1/2}$.
\salta

Let us introduce some notation. 
For the positive Bohr frequencies we shall write, see Eq. (\ref{21}),
\b\label{28}
\o_{mn}:= E_m-E_n = -\frac{e^2}{2a_0}\,\left( \frac{1}{m^2}-\frac{1}{n^2}\right)\,,
\quad  m>n\,,
\e
and for the matrix elements of the operators $D(k)$ we shall put
\b\label{29}
\begin{array}{rl}
\ds g_{mn}(k):= & \ds \< \psi_{m00},D(k)\psi_{n00}\>
\\[1.5ex]

= & \ds \<R_{m0}(|q|)Y_{00}(\theta,\phi), D(k)\,R_{m0}(|q|)Y_{00}(\theta,\phi)\>\,.
\end{array}
\e

Now suppose that for the positive Bohr frequency $\o_{mn}$ there exists a unique pair of energy levels $\vep_{m}=E_m$, $\vep_{n}=E_n$ in ${\rm Spec\,} H_S$ such that 
$\o_{mn}= E_m-E_n$.\footnote
{
This is not the case for every positive Bohr frequency $\o_{mn}$ of the hydrogen atom, but by means of a little perturbation the hydrogen atom becomes a system verifying this generic assumption.
} 
Then, the operator $D_{\o_{mn}}$ defined by Eq. (\ref{opD}) can be written as 
$$
D_{\o_{mn}}(k)= g_{mn}(k)\,|\psi_{n00}\>\< \psi_{m00}|
$$
and the imaginary part of the corresponding generalized susceptivity factor $\gamma_-^{\o_{mn}}$ is of the form,
see Eq. (\ref{alv4.21.3}),
\b\label{10}
Im(\gamma_-^{\o_{mn}})= -\pv \int_{\R^d} \frac{|g_{mn}(k)|^2}{\o(k)-\o}\, dk \,.
\e

In what follows we shall study the generalized Hilbert transform of the right hand side of Eq. (\ref{10}) in terms of the cutoff function $g$.
We restrict our attention to the dispersion function $\o(k)=|k|$ of example \ref{ex1}. For $\o(k)=k^2$ the results  are  very similar.

%--------------
\subsection{Cutoff Functions of the form $\ds \frac{1}{|k|^{\nu}}$, $(\nu\geq 0)$.}

When the response terms under consideration are of the form 
\b\label{20}
D(k)= \frac{e^{ik\cdot q}}{|k|^{\nu}}\,, \quad \nu\geq 0 \,,
\e  
the matrix element corresponding to the states of the electron with energy numbers $m$ and $n$ and orbital number $l=0$ is 
\b\label{35}
g_{mn}(k)= 
\frac{i}{2\,|k|^{1+\nu}}\,\sum_{s=2}^{m+n} C_s^{mn}\,\left(\frac{1}{(1+i|k|)^{s}}-\frac{1}{(1-i|k|)^{s}}\right)\,,
\e
where 
\b\label{34}
C_s^{mn}:= \frac{(-1)^{s}\, 4}{s\,a_0^3\,(mn)^{3/2}}\, 
\sum_{\alpha=\max\{0,s-m-1\}}^{\min\{n-1,s-2\}} \left(\begin{array}{c} n-1 \\ \alpha \end{array}\right) 
\left(\begin{array}{c} m-1 \\ s-2-\alpha \end{array}\right)
\left(\begin{array}{c} s \\ \alpha+1 \end{array}\right)\,.  
\e

The change of variables (\ref{chv}) gives us a more convenient expression of the right hand side of Eq. (\ref{10}).
For the dispersion function $\o(k)=|k|$ we have
\b\label{39x}
\begin{array}{r}
\ds  \pv \int_{\R^3} \frac{|g_{mn}(k)|^2}{\o(k)-\o_{mn}}\, dk = 
\ds  \pv \int_{-\o_{mn}}^\infty \frac{\pi}{u_1\,(u_1+\o_{mn})^{2\nu}}\times  
\\[2ex]
 \ds  \times \left|\sum_{s=2}^{m+n} C_s^{mn}
\left(\frac{1}{\big(1+i(u_1+\o_{mn})\big)^s}-\frac{1}{\big(1-i(u_1+\o_{mn})\big)^s}\right) \right|^2
\, du_1 \,.
\end{array}
\e

We can already determine the values of the parameter $\nu$ for which the imaginary part of the generalized susceptivity factor $\gamma_-^{\o_{mn}}$ exist.

\begin{prop}\label{prop9}
Let us consider the interaction of a spinless electron in the hydrogen atom with an electromagnetic field, in which the response terms  are of the form 
\b\label{20x}
D(k)= \frac{e^{ik\cdot q}}{|k|^{\nu}}\,, \quad \nu\geq 0 \,.
\e  
Then, for the dispersion functions $\o(k)=|k|$ (and also for $\o(k)=k^2$), the Cauchy Principal Values
\b\label{36x}
  \pv \int_{\R^3} \frac{|g_{mn}(k)|^2}{\o(k)-\o_{mn}}\, dk\,, \quad m>n\in\N\,,
\e
are finite if and only if $\nu<3/2$.
\end{prop}

%--------------
\subsection{Cutoff Functions of the form $\ds g(|k|)$.}

Now, let us consider response terms of the form 
\b\label{20c}
D(k)= g(|k|)\, e^{ik\cdot q}\,,
\e  
being the {\it cutoff function} $g$, at first, a medible function of $|k|$.

In this case the matrix element corresponding to the states of the electron with energy numbers $m$ and $n$ and orbital number $l=0$ is 
\b\label{35c}
g_{mn}(k)= 
\frac{i\,g(|k|)}{2\,|k|}\,\sum_{s=2}^{m+n} C_s^{mn}\,\left(\frac{1}{(1+i|k|)^{s}}-\frac{1}{(1-i|k|)^{s}}\right)\,,
\e
where the coefficient $C_s^{mn}$ is given by Eq. (\ref{34}).

For the dispersion function $\o(k)=|k|$, the change of variables (\ref{chv}) gives us now
\b\label{39xc}
\begin{array}{r}
\ds  \pv \int_{\R^3} \frac{|g_{mn}(k)|^2}{\o(k)-\o_{mn}}\, dk = 
\pv \int_{-\o_{mn}}^\infty \frac{\pi\,|g(u_1+\o_{mn})|^2}{u_1}\times
\\[2ex]
\ds   

\times \left|\sum_{s=2}^{m+n} C_s^{mn}\,
\left(\frac{1}{\big(1+i(u_1+\o_{mn})\big)^s}-\frac{1}{\big(1-i(u_1+\o_{mn})\big)^s}\right) \right|^2
\, du_1 \,.
\end{array}
\e

From the last expression we can deduce sufficient conditions on the form factor $g$ in order that the imaginary part of the generalized susceptivity factor $\gamma_-^{\o_{mn}}$ exist.

\begin{prop}\label{prop9c}
Let us consider the interaction of a spinless electron in the hydrogen atom with an electromagnetic field, in which the response terms  are of the form 
\b\label{20ccx}
D(k)= g(|k|)\, e^{ik\cdot q}\,,
\e  
Then, for the dispersion function $\o(k)=|k|$, the Cauchy Principal Values
\b\label{36xc}
  \pv \int_{\R^3} \frac{|g_{mn}(k)|^2}{\o(k)-\o_{mn}}\, dk\,, \quad m>n\in\N\,,
\e
are finite if the function $g$ verify the following conditions:
\begin{itemize}
\item[(a1)]
$|g(u_1+\o_{mn})|^2/u_1^4\in L^1\big([b,\infty)\big)$, with respect to $u_1$, for some $b>0$; for example, if $|g(u_1+\o_{mn})|\simeq u_1^\nu$ as $u_1\to\infty$, for $\nu<3/2$;
\item[(a2)]
the integrand of the right hand side of Eq. (\ref{39xc}), in a neighbourhood of $u_1=0$, is the sum of an antisymmetric function $f_1$ and a symmetric function $f_2$ such that $\int_{\to 0}^\epsilon f_2(u_1)\,du_1$ is finite for some $\epsilon>0$; for example, when $g$ is bounded in a neighbourhood of $\o_{mn}$ (or, equivalently, $g(u_1+\o_{mn})$ is bounded for $u_1$ in a neighbourhood of $0$);
\item[(a3)]
$|g(u_1+\o_{mn})|^2 (u_1+\o_{mn})^2\in L^1\big([-\o_{mn},-\o_{mn}+\epsilon]\big)$ with respect to $u_1$, for some $\epsilon<\o_{mn}$; for example, if $|g(u_1+\o_{mn})|\simeq (u_1+\o_{mn})^\nu$ as $u_1\to-\o_{mn}$, for $\nu>-3$;
\item[(a4)]
$g(u_1+\o_{mn})\in L^2\big([-\o_{mn}+\epsilon,-\epsilon]\cup [\epsilon,b]\big)$, with respect to $u_1$, for some $0<\epsilon<\o_{mn}$ and any finite $b>\epsilon$. 
\end{itemize}

\end{prop}

Note that condition (a2) implies some special behaviour of the cutoff function $g$ in a neighbourhood of the resonance surface $u_1=\o(k)-\o_{mn}=0$. 

\section*{Acknowledgements}
The author wishes to thank Prof. L. Accardi for useful discussions and the staff of Centro Interdepartamentale Vito Volterra of Universit\`a degli Studi di Roma ``Tor Vergata" for kind hospitality. 
This work was supported by JCyL-project UV95/02 (Castilla y Le\'on) and MEC-project BFM2002-02000 (Spain).


\begin{thebibliography}{99}

\bibitem{ALV}
L. Accardi, Y.G. Lu, I. Volovich, {\it Quantum Theory and Its Stochastic Limit}, Springer-Verlag, Berlin, 2002.

\bibitem{GS}
I.M. Gelfand, G.E. Shilov, {\it Les Distributions}, Dunod, Paris, 1962.

\bibitem{PW}
L. Pauling, E.B. Wilson Jr., {\it Introduction to Quantum Mechanics}, McGraw-Hill, New York, 1935.

\bibitem{SCH}
L. Schwartz, {\it M\'ethodes Math\'ematiques pour les Sciences Physiques}, Hermann, Paris, 1966.

\end{thebibliography}
\end{document}